\documentclass[12pt]{article}
\usepackage{amsmath}
\usepackage{graphicx}
\usepackage{enumerate}
\usepackage{natbib}
\usepackage{url}
\usepackage{hyperref}
\usepackage{gensymb}

\newcommand{\blind}{1}

\usepackage{comment}
\usepackage{amssymb}
\usepackage{mathtools}
\usepackage{multirow}
\usepackage{color}
\usepackage{bbm}
\usepackage{bm}
\usepackage{amsthm}

\newcommand{\bh}{\mathbf{h}}

\newcommand{\bs}{\mathbf{s}}

\newcommand{\bc}{\mathbf{c}}
\newcommand{\bC}{\mathbf{C}}

\newcommand{\bZ}{\mathbf{Z}}
\newcommand{\bSigma}{\mathbf{\Sigma}}

\newcommand{\bzero}{\bm{0}}

\newcommand{\cov}{\hbox{cov}}

\newcommand{\trans}{{^\top}}
\newcommand{\defeq}{\vcentcolon=}
\DeclarePairedDelimiter\abs{\lvert}{\rvert}%

\newtheorem{theorem}{Theorem}[section] 
\newtheorem{proposition}[theorem]{Proposition} 
\newtheorem{model}[theorem]{Model}

\begin{document}

\def\spacingset#1{\renewcommand{\baselinestretch}%
	{#1}\small\normalsize} \spacingset{1}

\if1\blind
{
	\title{\bf
		Test and Visualization of Covariance Properties for Multivariate Spatio-Temporal Random Fields}
	\author{Huang Huang\thanks{huang.huang@kaust.edu.sa}\hfill
		Ying Sun\hfill
		Marc G. Genton\\
		Statistics Program\\
		King Abdullah University of Science and Technology\\
	}
	\maketitle
} \fi

\if0\blind
{
	\bigskip
	\bigskip
	\bigskip
	\begin{center}
		{\LARGE\bf }
	\end{center}
	\medskip
} \fi

\begin{center}
Published in  \textit{Journal of Computational and Graphical Statistics}
\end{center}

\bigskip

\begin{abstract}
The prevalence of multivariate space-time data collected from monitoring networks and satellites, or generated from numerical models, has brought much attention to multivariate spatio-temporal statistical models, where the covariance function plays a key role in modeling, inference, and prediction.
For multivariate space-time data, understanding the spatio-temporal variability, within and across variables, is essential in employing a realistic covariance model.
Meanwhile, the complexity of generic covariances often makes model fitting very challenging, and simplified covariance structures, including symmetry and separability, can reduce the model complexity and facilitate the inference procedure. 
However, a careful examination of these properties is needed in real applications.
In the work presented here, we formally define these properties for multivariate spatio-temporal random fields and use functional data analysis techniques to visualize them, hence providing intuitive interpretations. 
We then propose a rigorous rank-based testing procedure to conclude whether the simplified properties of covariance are suitable for the underlying multivariate space-time data.
The good performance of our method is illustrated through synthetic data, for which we know the true structure. We also investigate the covariance of bivariate wind speed, a key variable in renewable energy, over a coastal and an inland area in Saudi Arabia.
The Supplementary Material is available online, including the \textsf{R} code for our developed methods.
\end{abstract}

\noindent%
\vfill

\newpage
\spacingset{1.5} 

\section{Introduction}
Multivariate spatio-temporal modeling has become a very active research area in many scientific fields, including climate, hydrology, and ecology, due to its capacity to capture the space-time characteristics of data and provide accurate statistical inference.
For example, \citet{PM2009} used multivariate spatio-temporal models to uncover forest composition in history based on fossil pollen records; \citet{ZRSLB2015} used a multivariate spatio-temporal Gaussian Markov random field to assess Antarctica's mass balance and contribution to the rise of sea-level;
\citet{MLPCTGB2019} used a multivariate spatio-temporal model in a Bayesian hierarchical framework to investigate the extreme temperatures and precipitation jointly.
Gaussian random fields are widely used in geostatistics, directly representing the variables used in the study or serving as a building block in more complex statistical models, where the covariance structure plays a key role in quantifying dependence and providing prediction.
\citet{GKS2010} developed valid Mat\'ern covariance functions for multivariate spatial random fields. \citet{AG2010} proposed new approaches to build valid multivariate spatio-temporal covariance models via latent dimensions. \citet{GK2015} reviewed approaches to build cross-covariance functions for multivariate random fields.

For real applications, the choice of the covariance model is case-specific. Understanding the spatio-temporal variability within and across variables is essential in choosing a realistic covariance model. In addition, with the development of remote sensing and \textit{in situ} measurement techniques, and with powerful computing facilities that enable better physical model simulations, large data sets of unprecedented size are collected.
It is often challenging and slow to perform model inference because space-time covariance models typically involve many parameters that need to be estimated from data. Specific constraints on the proposed covariance structure can reduce the model complexity and accelerate the inference procedure.
These considerations motivate researchers to study simplified covariance structures and propose tools to visualize and assess them.
\citet{CH1999} proposed several approaches to build univariate nonseparable spatio-temporal covariances with separable covariances as special cases.
\citet{GGG2007} investigated both the univariate separability and symmetry covariance structures.
Graphical evidence for these univariate properties includes the contour plots in \citet{CH1999} and the functional boxplot \citep{SG2011} of proposed test functions in \citet{HS2019}.
Formal testing approaches have also been developed for the assessment of these properties. Examples include the test based on spectral representations~\citep{F2006} and the likelihood ratio test~\citep{MGG2006} for the univariate separability property, the separability and symmetry tests from constructed contrasts of convariances~\citep{LGS2007}, and the rank-based testing procedure using functional data analysis techniques~\citep{HS2019};
see \citet{CGS2021} for a comprehensive review of univariate spatio-temporal covariance structures and models. 
When data are multivariate space-time, there are more types of covariance structures, and the assessment is more complicated.
\citet{W2013} used graphical evidence to indicate multivariate separability and symmetry. 
\citet{LGS2008} extended their previous testing procedure of the univariate covariance to the multivariate case, where the property is assessed at selected spatio-temporal lags by examining the test statistic built from the asymptotic distribution. 

To the best of our knowledge, the description of the different types of multivariate spatio-temporal separability and symmetry properties is scattered in the literature and incomplete, and there has not been any work formally defining all the different types of multivariate spatio-temporal covariances. In this paper, we introduce several possible types of separability and symmetry properties for the multivariate spatio-temporal covariance and give formal definitions.
We then develop test functions associated with each of the properties and use functional data analysis techniques to visualize them. Our proposed visualization tool is a very fast approach to view these properties; there has not been such a tool developed yet in the multivariate setting. We also propose a rank-based testing procedure that can examine these properties quantitatively.
For simplicity, we focus on a multivariate strictly stationary spatio-temporal random field $\bZ(\bs,t)\in\mathbb{R}^p$ for location $\bs\in\mathcal{D}\subset\mathbb{R}^d$ and time $t\in \{1, \ldots, l\}$, where the $p\times p$ matrix-valued stationary covariance function $\bC(\bh,u) = \cov\{\bZ(\bs_1,t_1),\bZ(\bs_2,t_2)\}$ depends only on the space lag $\bh = \bs_2-\bs_1$ and time lag $u = t_2 - t_1$.

Our paper is organized as follows: Section~\ref{sec:properties} defines multivariate spatio-temporal covariance properties to be studied and our proposed tools for property visualization and assessment;
Section~\ref{sec:methodology} describes our new methodology to visualize our proposed test functions and formally test covariance properties;
Section~\ref{sec:sim} uses synthetic data sets to demonstrate the performance of the proposed testing procedure;
Section~\ref{sec:app} provides an example where we apply our proposed tools to analyze bivariate hourly wind speed over a coastal and an inland area in Saudi Arabia;
Section~\ref{sec:discussion} summarizes our methods and presents research directions for future improvement.

\section{Multivariate Spatio-Temporal Covariance Properties}\label{sec:properties}
For a $p$-variate strictly stationary spatio-temporal random field $\bZ(\bs,t) = \{Z_1(\bs,t),\ldots,$ $ Z_p(\bs,t) \}\trans$, we denote the covariance between the $i$-th and $j$-th variable with space lag $\bh$ and time lag $u$ as $C_{ij}(\bh,u)=\cov\{Z_i(\bs+\bh,t+u),Z_j(\bs,t)\}$, $i,j=1,\ldots,p, \bs\in\mathcal{D}, \bs+\bh\in\mathcal{D}.$  
\subsection{Symmetry structures}
For a univariate spatio-temporal covariance, there is a unique symmetry property; symmetry means $C(\bh,u)=C(\bh,-u)$, equivalent to $C(\bh,u)=C(-\bh,u)$, where $C(\bh,u) = \cov\{Z(\bs+\bh,t+u),Z(\bs,t)\}$ for a univariate spatio-temporal random field $Z(\bs,t)$ \citep{HS2019}. 
When extending it to the multivariate case, more types of symmetry may occur.
The simplest case is the \textit{fully symmetric} covariance, where $C_{ij}(\bh,u) = C_{ji}(\bh,u) = C_{ij}(-\bh,u) = C_{ij}(\bh,-u), \forall i,j\in\{1,\ldots,p\}$.
When the covariance does not meet all the equality requirements, we further define more symmetry cases where only part of the equality requirements holds.
We call the covariance \textit{symmetric in variables} (denoted by $\text{V}^\text{sym}$) if $C_{ij}(\bh,u) = C_{ji}(\bh,u), \forall i,j\in\{1,\ldots,p\}$. This condition is equivalent to $C_{ij}(\bh,u) = C_{ij}(-\bh,-u)$ due to the fact that $C_{ij}(\bh,u) = C_{ji}(-\bh,-u)$ naturally holds.
We call the covariance \textit{symmetric in space} (denoted by $\text{S}^\text{sym}$) if $C_{ij}(\bh,u) = C_{ij}(-\bh,u), \forall i,j\in\{1,\ldots,p\}$ and similarly
\textit{symmetric in time} (denoted by $\text{T}^\text{sym}$) if $C_{ij}(\bh,u) = C_{ij}(\bh,-u), \forall i,j\in\{1,\ldots,p\}$.
The properties of symmetry in variables, space, and time are not independent; Proposition~\ref{prop:sym} shows the relationship among them with the proof given in the Supplementary Material.

\begin{proposition}\label{prop:sym} If a multivariate spatio-temporal stationary covariance has any two properties among 1) symmetry in variables, 2) symmetry in space, or 3) symmetry in time, then the covariance is fully symmetric, which we denote by  denoted by $\text{F}^\text{sym}$.
\end{proposition}

\subsection{Separability structures}
Several types of separability can happen for multivariate spatio-temporal random fields, and we define them in Table~\ref{tab:separability-definition}, where we also show the relation of symmetry properties if there is one. 
Some abbreviated notations for ease of presentation are introduced as follows:
V$\mid$ST, \textit{separability between variables and space-time};
S$\mid$VT, \textit{separability between space and variables-time};
T$\mid$VS, \textit{separability between time and variables-space};
V$\mid$S, \textit{separability between variables and space};
V$\mid$T, \textit{separability between variables and time};
S$\mid$T, \textit{separability between space and time};

\begin{table}[ht!]
	\centering
	\caption{Definition of different types of separability properties. For each type, the covariance function can be written as two components, one of which is a general function concerning only a part of variables, space, and time, denoted by $\rho$. For example, $\rho_1(\bh,u)$ is a function with respect to only space and time. There are some natural equalities for these functions $\rho$: all these functions equal 1 when $\bh=\bzero$ and $u=0$; $\rho_4(\bzero,u)= 1$ for any $u$; and $\rho_5(\bh,0)= 1$ for any $\bh$.
	}
	\begin{tabular}{c|c|c}
		\hline
		\textbf{Type} & \textbf{Definition}  & \textbf{Implication} \\
		\hline
		
		{V$\mid$ST} & 
		$C_{ij}(\bh,u) =  \rho_1(\bh,u)C_{ij}(\bzero,0), \forall i,j\in\{1,\ldots,p\}$ & 
		$\text{V}^\text{sym}$
		\\
		\hline
		
		{S$\mid$VT} & 
		$C_{ij}(\bh,u) = \rho_2(\bh)C_{ij}(\bzero,u), \forall i,j\in\{1,\ldots,p\}$ & 
		$\text{S}^\text{sym}$
		\\
		
		\hline
		
		{T$\mid$VS} & 
		$C_{ij}(\bh,u) = \rho_3(u) C_{ij}(\bh,0), \forall i,j\in\{1,\ldots,p\}$ & 
		$\text{T}^\text{sym}$
		\\
		
		\hline
		
		{V$\mid$S} & 
		$C_{ij}(\bh,u) = \rho_4(\bh,u)C_{ij}(\bzero,u), \forall i,j\in\{1,\ldots,p\}$ &
		\\
		
		\hline
		
		{V$\mid$T} & 
		$C_{ij}(\bh,u) = \rho_5(\bh,u)C_{ij}(\bh,0), \forall i,j\in\{1,\ldots,p\}$&
		\\
		
		\hline
		
		{S$\mid$T} & 
		$C_{ij}(\bh,u) = \rho_{6,ij}(\bh) C_{ij}(\bzero,u), \forall i,j\in\{1,\ldots,p\}$ & 
		\\
		
		\hline
		
	\end{tabular}
	\label{tab:separability-definition}
\end{table}

We observe that the extension to the multivariate case introduces many more types of separability than the single separability property between space and time for univariate spatio-temporal random fields.
One can clearly see that these six types can be grouped into two categories.
The first category is that one component is completely separated from the other two. Though this may rarely happen for real data, the benefit brought by this property is significant because it allows us to write the covariance matrix as a Kronecker product of two smaller matrices.
The three cases in this category are V$\mid$ST, S$\mid$VT, and T$\mid$VS.
For example, when the covariance is V$\mid$ST, it can be decomposed into two parts corresponding to the covariance among variables ($C_{ij}(\bzero,0)$) and the spatio-temporal correlation ($\rho_1(\bh,u)$). This is also known as the \textit{intrinsic correlation model} in \cite{W2013}. We know $C_{ij}(\bh,u) =  \rho_1(\bh,u)C_{ij}(\bzero,0) =  \rho_1(\bh,u)C_{ji}(\bzero,0) = C_{ji}(\bh,u)$, so the multivariate spatio-temporal covariance is also $\text{V}^\text{sym}$.

The second category is the one for which no component in the covariance can be completely separated, but the interaction between certain two out of the three components is voided, which has less restriction than the first category described above. The three cases in the second category are V$\mid$S, V$\mid$T, and S$\mid$T.
For example, when the covariance is V$\mid$S, there is no variable-space interaction, and the covariance can be decomposed into two parts corresponding to the variable-temporal correlation ($C_{ij}(\bzero,u)$) and spatio-temporal correlation ($\rho_4(\bh,u)$).
For cases in the second category, no symmetry property holds.

We also call a multivariate spatio-temporal covariance \textit{fully separable} (denoted by $\text{F}^\text{sep}$) if it satisfies all the separability properties, which further implies $\text{F}^\text{sym}$.
Like for symmetry, there are some constraints among different types of separability summarized in Propositions~\ref{prop:sep1} through \ref{prop:sep3}, proofs of which are given in the Supplementary Material.

\begin{proposition}\label{prop:sep1} If a covariance is \textup{V}$\mid$\textup{ST}, then it is naturally \textup{V}$\mid$\textup{S} and \textup{V}$\mid$\textup{T}. Similarly, \textup{S}$\mid$\textup{VT} implies \textup{V}$\mid$\textup{S} and \textup{S}$\mid$\textup{T}, and \textup{T}$\mid$\textup{VS} implies \textup{V}$\mid$\textup{T} and \textup{S}$\mid$\textup{T}.
\end{proposition}

\begin{proposition}\label{prop:sep2} If a covariance is \textup{V}$\mid$\textup{S} and \textup{V}$\mid$\textup{T}, then the covariance is \textup{V}$\mid$\textup{ST}. Similarly, \textup{V}$\mid$\textup{S} and \textup{S}$\mid$\textup{T} imply \textup{S}$\mid$\textup{VT}, and  \textup{V}$\mid$\textup{T} and \textup{S}$\mid$\textup{T} imply \textup{T}$\mid$\textup{VS}.
\end{proposition}

\begin{proposition}\label{prop:sep3} If any two properties of  \textup{V}$\mid$\textup{ST}, \textup{S}$\mid$\textup{VT}, or \textup{T}$\mid$\textup{VS} hold, then the remaining one also holds and the covariance is $\textup{F}^\textup{sep}$.
\end{proposition}

Note that in this work, we aim to study the overall property of the multivariate spatio-temporal covariance. Indeed, the properties are assumed to hold uniformly across all space lags $\bh$ and time lags $u$, which is different from the point-wise manner of \citet{DPM2013}.

\section{Methodology\label{sec:methodology}}
\subsection{Test functions\label{subsec:tf}}

The essential idea of examining different covariance properties is to propose associated test functions whose mean is zero when the property holds.
Our test functions are a collection of realized random functions of temporal lag $u$ for every pair of variables and locations, defined through the sample estimate of 
$C^{a,b}_{ij}(\bs_b-\bs_a,u) = \cov \big\{Z_i(\bs_a,t), Z_j(\bs_b,t+u)\big\}$
for $i,j\in\{1,\ldots,p\}$, $t \in \{1, \ldots, l-u\}$, and locations $\bs_a,\bs_b\in\mathcal{D}$, where $l$ is the number of time points, as follows:
\[   
\hat C^{a,b}_{ij}(\bs_b-\bs_a,u) \defeq \dfrac{1}{l-u}
\sum^{l-u}_{t=1}
\Bigg\{Z_j(\bs_b,t+u)-\frac{\sum\limits^{l-u}_{r=1}Z_j(\bs_b,r+u)}{l-u}\Bigg\}
\Bigg\{Z_i(\bs_a,t)-\frac{\sum\limits^{l-u}_{r=1}Z_i(\bs_a,r)}{l-u}\Bigg\}.
\]
\subsubsection{Test functions for Symmetry \label{subsubsec:tfsym}}

Table~\ref{tab:symmetry-test-func} lists the definitions of the test functions examining each symmetry property, where the indices of $i,j,a$, and $b$ are selected such that trivial zero curves and redundant realizations are not considered.
For example, the test functions for $\text{V}^\text{sym}$ are the differences between the sample estimators of covariance at particular lags when the order of variables is flipped.
It is easy to observe that the expectation of the test functions is zero when the corresponding symmetry property holds and vice versa. 
For $p$-variate spatio-temporal random fields at $n$ locations, the number of test function realizations for $\text{V}^\text{sym}$, $\text{S}^\text{sym}$, and $\text{T}^\text{sym}$ are $n^2p(p-1)/2, n(n-1)p^2/2$, and $n^2p(p+1)/2-np$, respectively.

\begin{table}[htp!]
	\centering
	\caption{Definition of test functions for different types of symmetry.}
	\begin{tabular}{c|c}
		\hline
		\textbf{Type} & \textbf{Test Functions} \\
		\hline
		
		$\text{V}^\text{sym}$
		& 
		$  g^v_{i,j,a,b}(u)  \defeq  \hat C^{a,b}_{ij}(\bs_b-\bs_a,u) - \hat C^{a,b}_{ji}(\bs_b-\bs_a,u), u\in\mathbb{N}, i<j$\\
		\hline
		
		$\text{S}^\text{sym}$
		& 
		$g^s_{i,j,a,b}(u) \defeq \hat C^{a,b}_{ij}(\bs_b-\bs_a,u) - \hat C^{b,a}_{ij}(\bs_a-\bs_b,u), u\in\mathbb{N};
		i\leq j, a < b \text{ or } i > j, a > b$\\
		\hline
		
		$\text{T}^\text{sym}$
		&
		$g^t_{i,j,a,b}(u) \defeq \hat C^{a,b}_{ij}(\bs_b-\bs_a,u) - \hat C^{a,b}_{ij}(\bs_b-\bs_a,-u), u\in\mathbb{N}_+;
		i< j \text{ or } i = j, a\neq b$\\
		\hline
		
	\end{tabular}
	
	\label{tab:symmetry-test-func}
\end{table}

\subsubsection{Test functions for Separability \label{subsubsec:tfsep}}
To build test functions for separability, in addition to estimating $C_{ij}(\bh,u)$, we also need to estimate all the functions $\rho_1(\bh,u)$, $\rho_2(\bh)$, $\rho_3(u)$, $\rho_4(\bh,u)$, $\rho_5(\bh,u)$, and $\rho_{6,ij}(\bh)$. 
Various approaches are possible to estimate them. 
For example, $\hat \rho_1^{a,b}(\bs_b-\bs_a,u)$ can be $\sum^p_{i,j=1}[2\hat C^{a,b}_{ij}(\bs_a-\bs_b,u)/\{\hat C^{a,a}_{ij}(\bzero,u)+\hat C^{b,b}_{ij}(\bzero,u)\}]/p$, 
which is the average of the estimator of $\rho_1^{a,b}(\bs_a-\bs_b,u)$ from $Z_i(\bs_a,t)$ and $Z_j(\bs_b,t)$ for each $i,j\in\{1,\ldots,p\}$ (we call this a mean-ratio estimator).
To obtain the best linear unbiased estimate, we regress $\hat C^{a,b}_{ij}(\bs_a-\bs_b,u)$ on $\{\hat C^{a,a}_{ij}(\bzero,u)+\hat C^{b,b}_{ij}(\bzero,u)\}/2$ and get the least-square estimator of $\rho_1^{a,b}(\bs_a-\bs_b,u)$ as follows:
\[
\hat\rho_1^{a,b}(\bs_a-\bs_b,u)
=
\dfrac{2\sum^{p}_{i,j=1}\hat C^{a,b}_{ij}(\bs_a-\bs_b,u)\{\hat C^{a,a}_{ij}(\bzero,0)+\hat C^{b,b}_{ij}(\bzero,0)\}}
{\sum^{p}_{i,j=1}\{\hat C^{a,a}_{ij}(\bzero,0)+\hat C^{b,b}_{ij}(\bzero,0)\}^2},
\]
where we assume ${\sum^{p}_{i,j=1}\{\hat C^{a,a}_{ij}(\bzero,0)+\hat C^{b,b}_{ij}(\bzero,0)\}^2}\neq0$.
The least-square estimators for all the other required functions are as follows, where we also assume the associated denominators are not zero in each case:
\[
\begin{array}{rcl}

\hat\rho_2^{a,b}(\bs_a-\bs_b)
&=&
\dfrac{2\sum^{p}_{i,j=1}\hat C^{a,b}_{ij}(\bs_a-\bs_b,0)\{\hat C^{a,a}_{ij}(\bzero,0)+\hat C^{b,b}_{ij}(\bzero,0)\}}
{\sum^{p}_{i,j=1}\{\hat C^{a,a}_{ij}(\bzero,0)+\hat C^{b,b}_{ij}(\bzero,0)\}^2},\\

\hat\rho^{a,b}_{3}(u)
&=&
\dfrac{\sum^{p}_{i,j=1}\{\hat C^{a,a}_{ij}(\bzero,u)+\hat C^{b,b}_{ij}(\bzero,u)\}\{\hat C^{a,a}_{ij}(\bzero,0)+\hat C^{b,b}_{ij}(\bzero,0)\}}
{\sum^{p}_{i,j=1}\{\hat C^{a,a}_{ij}(\bzero,0)+\hat C^{b,b}_{ij}(\bzero,0)\}^2},\\

\hat{\rho}^{a,b}_4(\bs_a-\bs_b,u)
&=&
\dfrac{2\sum^{p}_{i,j=1}\hat C^{a,b}_{ij}(\bs_a-\bs_b,u)\{\hat C^{a,a}_{ij}(\bzero,u)+\hat C^{b,b}_{ij}(\bzero,u)\}}
{\sum^{p}_{i,j=1}\{\hat C^{a,a}_{ij}(\bzero,u)+\hat C^{b,b}_{ij}(\bzero,u)\}^2},\\

\hat{\rho}^{a,b}_5(\bs_a-\bs_b,u)
&=&
\dfrac{\sum^{p}_{i,j=1}\hat C^{a,b}_{ij}(\bs_a-\bs_b,u)\hat C^{a,b}_{ij}(\bs_a-\bs_b,0)}
{\sum^{p}_{i,j=1}\hat C^{a,b}_{ij}(\bs_a-\bs_b,0)^2},\\

\hat\rho^{a,b}_{6,ij}(\bs_a-\bs_b)
&=&
{2\hat C^{a,b}_{ij}(\bs_a-\bs_b,0)}/
{\{\hat C^{a,a}_{ij}(\bzero,0)+\hat C^{b,b}_{ij}(\bzero,0)\}}.\\

\end{array}
\]

With estimators of these $\rho$-functions, we can define the test functions for different types of separability, as shown in Table~\ref{tab:separability-test-func}, where we also remove trivial zero test functions.
For example, the test functions for V$\mid$ST are the differences between the estimates of covariance and the product of covariance among variables and spatio-temporal correlation at particular lags.
For $p$-variate spatio-temporal random fields at $n$ locations, the numbers of test function realizations are $n^2p^2$ for V$\mid$ST, T$\mid$VS, V$\mid$S, and V$\mid$T, and $n(n-1)p^2$ for S$\mid$VT  and S$\mid$T.
Unlike for symmetry cases, we only have asymptotic results for the expectation of the test functions under some mild conditions, as given in Theorem~\ref{th:sep} with proof provided in the Supplementary Material.

\begin{theorem}
	\label{th:sep}
	If the covariance function $C_{ij}(\bh,u)$ for a multivariate strictly stationary spatio-temporal random field $\bZ(\bs,t)=\{Z_1(\bs,t),\ldots,Z_p(\bs,t)\}^\top$ satisfies the conditions:
	\begin{enumerate}
		\item $\sum\limits_{t\in\mathbb{Z}}
		\abs{\cov\{Z_i(\bs_a,0)Z_{j}(\bs_b,u_1),Z_{i'}(\bs_{a'},t)Z_{j'}(\bs_{b'},t+u_2)\}}<\infty$, for any finite $u_1,u_2\in\mathbb{Z}$, $i, j, i', j' \in\{1,\ldots,p\}$, and $\bs_a,\bs_b,\bs_{a'},\bs_{b'}\in\mathcal{D}$,
		\item $\sum\limits^{p}_{i,j=1} C^2_{ij}(\bzero,u)\neq 0$ for the case of separability between variable and space, and $\sum\limits^{p}_{i,j=1}C^2_{ij}(\bzero,0)\neq 0$ for all other separability cases,
	\end{enumerate}
	 then the expectation of the test functions converges to zero as the number of time points goes to infinity when the covariance function is separable with the corresponding type.
\end{theorem}

\begin{table}[ht!]
	\centering
	\caption{Definition of test functions for different types of separability.}
	\begin{tabular}{c|c}
		\hline
		\textbf{Type} & \textbf{Test Functions} \\
		\hline
		
		V$\mid$ST& 
		$
		f^{v\mid st}_{i,j,a,b}(u) \defeq
		\hat C^{a,b}_{ij}(\bs_b-\bs_a,u) - \hat \rho^{a,b}_1(\bs_b-\bs_a,u)\{\hat C^{a,a}_{ij}(\bzero,0)+\hat C^{b,b}_{ij}(\bzero,0)\}/2,
		u\in\mathbb{N}_+	
		$ \\
		\hline
		
		S$\mid$VT& 
		$
		f^{s\mid vt}_{i,j,a,b}(u) \defeq
		\hat C^{a,b}_{ij}(\bs_b-\bs_a,u) - \hat \rho^{a,b}_2(\bs_b-\bs_a)\{\hat C^{a,a}_{ij}(\bzero,u)+\hat C^{b,b}_{ij}(\bzero,u)\}/2,
		u\in\mathbb{N}_+, a\neq b
		$ \\
		\hline
		
		T$\mid$VS& 
		$
		f^{t\mid vs}_{i,j,a,b}(u) \defeq 
		\hat C^{a,b}_{ij}(\bs_b-\bs_a,u) - \hat \rho^{a,b}_3(u)\hat C^{a,b}_{ij}(\bs_b-\bs_a,0), u\in\mathbb{N}_+
		$ \\
		
		\hline
		
		V$\mid$S& 
		$
		f^{v\mid s}_{i,j,a,b}(u) \defeq
		\hat C^{a,b}_{ij}(\bs_b-\bs_a,u) - \hat \rho^{a,b}_4(\bs_b-\bs_a,u)\{\hat C^{a,a}_{ij}(\bzero,u)+\hat C^{b,b}_{ij}(\bzero,u)\}/2, u\in\mathbb{N}_+
		$ \\
		
		\hline
		
		V$\mid$T& 
		$
		f^{v\mid t}_{i,j,a,b}(u) \defeq
		\hat C^{a,b}_{ij}(\bs_b-\bs_a,u) - \hat \rho^{a,b}_5(\bs_b-\bs_a,u)\hat C^{a,b}_{ij}(\bs_b-\bs_a,0), u\in\mathbb{N}_+
		$ \\
		
		\hline
		
		S$\mid$T& 
		$
		f^{s\mid t}_{i,j,a,b}(u) \defeq
		\hat C^{a,b}_{ij}(\bs_b-\bs_a,u) - \hat \rho^{a,b}_{6,ij}(\bs_b-\bs_a)\{\hat C^{a,a}_{ij}(\bzero,u)+\hat C^{b,b}_{ij}(\bzero,u)\}/2
		,
		u\in\mathbb{N}_+,a\neq b
		$ \\
		
		\hline
		
	\end{tabular}
	\label{tab:separability-test-func}
\end{table}

\subsection{Visualization of test functions}
We see that, under the symmetry or separability assumption, the expectation of the associated test functions is zero or asymptotically zero. 
The deviation of test functions from zero suggests that the underlying assumption is violated.
To visually summarize the set of test functions for given data, we use functional boxplots~\citep{SG2011} with particular modification and extension to emphasize the deviation from zero.
In our visualization tool, we first compute the test functions and order them by the modified band depth~\citep{LR2009}.
The modified band depth is one type of functional data depths, the computation of which is based on the graphical representation of the functional data. It shows great computational efficiency and statistical power in detecting outliers and identifying the representative realizations.
We build the $50\%$ central region by the $50\%$ test functions with the largest band depth values, the border of which is drawn in blue.
To better show how these representative test functions are distributed in the central region, we compute the density of test functions falling into each area in the central region and fill in each area with a color whose opacity is proportional to the density.
A horizontal black dotted line is drawn to indicate zero.
At each $u$, we move the upper (lower) border of the central region upwards (downwards) by 1.5 times the range of the $50\%$ central region and get the outlier thresholds; the test functions that lie beyond the thresholds at any $u$ are detected as outliers. The whiskers, which are the envelope of the remaining test functions, are also drawn in blue.

\begin{figure}[ht!]
	\centering
	\includegraphics[width=\textwidth]{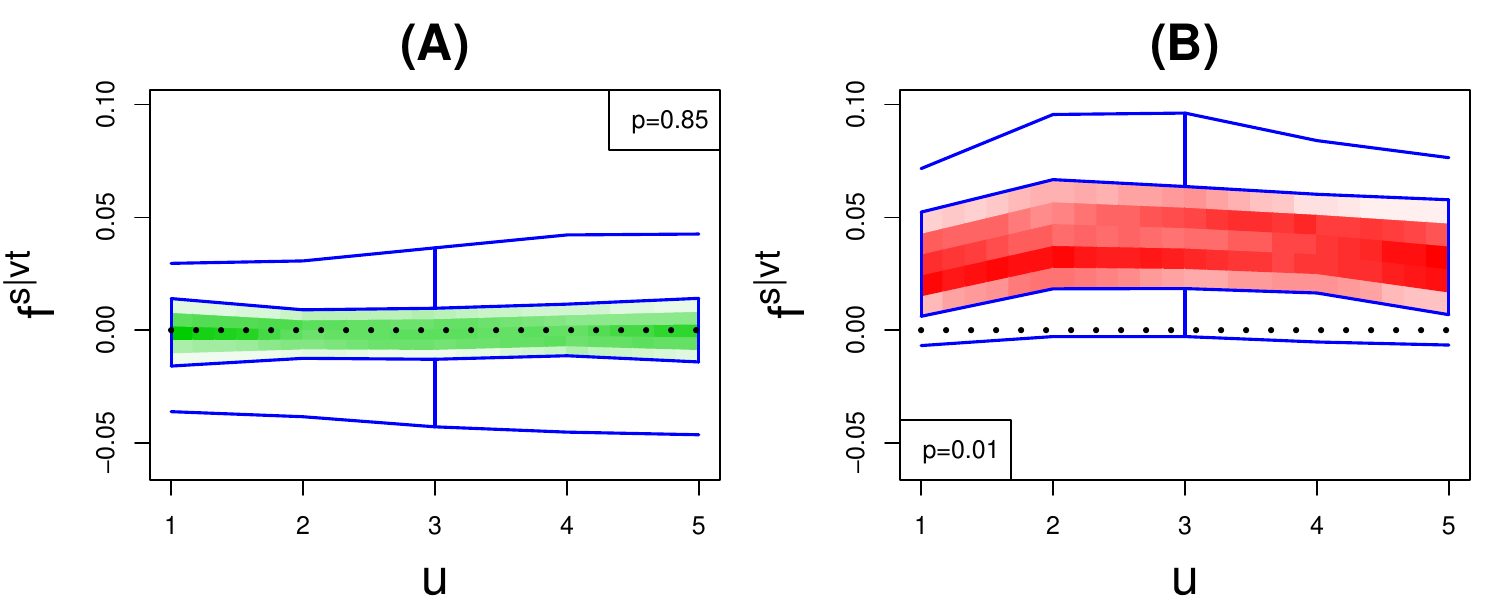}
	\caption{
		Visualization of test functions $f^{s\mid vt}$ for data generated with  $\beta_1=\beta_2=0$ (A) and $\beta_1=\beta_2=1$ (B) from Model~\ref{model:sep} in Section~\ref{model:sep}.
		P-values of the S$\mid$VT test  for the two data sets are shown in each panel. The central region is filled with green (as the p-value is greater than 0.05) and red (as the p-value is less than 0.05) for (A) and (B), respectively, whose opacity is proportional to the density of test functions falling into each area in the central region. The horizontal black dotted line indicates zero. The upper and lower blue curves are the whiskers.}
	\label{fig:tfplot_example}
\end{figure}

Figure~\ref{fig:tfplot_example} shows examples of our visualization tools of test functions $f^{s\mid vt}$ for data generated from Model~\ref{model:sep} detailed in Section~\ref{model:sep}.
Figure~\ref{fig:tfplot_example} (A) and (B) illustrate our modified functional boxplot of $f^{s\mid vt}$ for data simulated with $\beta_1=\beta_2=0$ (S$\mid$VT) and $\beta_1=\beta_2=1$ (not S$\mid$VT), respectively. 
The visualization tool also applies the testing procedure explained in Section~\ref{subsec:testing} and obtains the conclusion that the covariance is S$\mid$VT for Figure~\ref{fig:tfplot_example} (A) and not S$\mid$VT for  (B), with p-values shown in the plot. 
Therefore, green is used to fill in the central region in Figure~\ref{fig:tfplot_example} (A), meaning that the null hypothesis is not rejected with the significance level of $5\%$, and red is used in (B), meaning that the null hypothesis is rejected.
We observe that the central region in Figure~\ref{fig:tfplot_example} (A) tends to be symmetric around zero (black dotted line) due to estimation noise, but the central region in (B) is entirely above zero.

\subsection{Testing covariance properties\label{subsec:testing}}
For multivariate spatio-temporal Gaussian random fields, after obtaining the test functions, we can perform hypothesis tests for covariance properties based on nonparametric functional data ranking.
We use a similar testing procedure to that introduced by \citet{HS2019} for a univariate spatio-temporal covariance.
The key idea is to test whether two functional data sets are from the same distribution~\citep{LR2009}.
For ease of presentation, we denote by $F$ the collection of test functions for an arbitrary property. Then, we denote by $H_0$ and  $H_a$ the null and alternative hypotheses, respectively. For example, when we test $\text{V}^\text{sym}$, $F=\{g^v_{i,j,a,b}(u): i,j\in\{1,\ldots,p\}, \bs_s,\bs_b\in\mathcal{D}$\}, $H_0$ is symmetry in variables, and $H_a$ is asymmetry in variables.
We summarize the hypothesis test procedure as follows:
\begin{itemize}
	\item Step 1. We compute all the test functions from the data and obtain $F$.
	\item Step 2. We build the covariance matrix $\hat{\mathbf{C}}^{H_0}(\bh,u) = \{ \hat C_{ij}^{H_0}(\bh,u)\}_{i,j=1,\ldots,p}$ under $H_0$ with forms given in Table~\ref{tab:constructed-cov}, where $\hat C_{ij}(\bh,u)$, $\hat\rho_1(\bh,u)$, $\hat\rho_2(\bh)$, $\hat\rho_3(u)$, $\hat\rho_4(\bh,u)$, $\hat\rho_5(\bh,u)$, or $\hat\rho_{6,ij}(\bh)$ are the sample estimators from the data.
	\item Step 3. The obtained covariance matrix $\hat{\mathbf{C}}^{H_0}(\bh,u)$ may not necessarily be positive definite due to estimation noise. We find the nearest positive definite matrix to $\hat{\mathbf{C}}^{H_0}(\bh,u)$ in the Frobenius norm, using the function 
	\texttt{nearPD} from the \textsf{R}~\citep{R2021} package \texttt{Matrix}. Then, we generate two independent reference data sets with this covariance matrix and of the same dimension as the original data set.
	\item Step 4. We calculate the test function collection from these two reference data sets, denoted by $F_1^{H_0}$ and $F_2^{H_0}$,
	both obtained from the simulated data samples with the same covariance under $H_0$. 
	\item Step 5. Because the test functions in $F_1^{H_0}$ and $F_2^{H_0}$ should be close to zero, we apply the rank-based test for $F$ versus $F_1^{H_0}$ with the reference $F_2^{H_0}$. More precisely, suppose that there are $n_F$ and $n_{F_1^{H_0}}$ curves in $F$ and $F_1^{H_0}$, respectively. For each test function in $F$, we combine it with all test functions in $F_2^{H_0}$, and calculate its rank using an increasing order of modified band depths~\citep{LR2009}, denoted by $r_1,r_2,\ldots,r_{n_F}$.
	We do the same for each test function in $F_1^{H_0}$ and obtain the rank among $F_2^{H_0}$, denoted by $r'_1,r'_2,\ldots,r'_{n_{F_1^{H_0}}}$. 
	\item Step 6. We calculate the ranks of $r_1,r_2,\ldots,r_{n_F}$ in $\{r_1,r_2,\ldots,r_{n_F}, r'_1,r'_2,\ldots,r'_{n_{F_1^{H_0}}}\}$ in an increasing order and denote them by $q_1,q_2,\ldots,q_{n_F}$. The final test statistic is $W=\sum^{n_F}_{i=1}q_i.$
	The limiting distribution of $W$ under the null hypothesis $H_0$ is the sum of $n_F$ random samples from the integer sequence $1,\ldots,n_F+n_{F_1^{H_0}}$ without replacement~\citep{LS1993}. 
	The null hypothesis $H_0$ is rejected when $W$ is small.
	We can use the limiting distribution to obtain the p-values. However, in practice, we observe that when we apply the test to a simulated synthetic data set under $H_0$ many times, the resulting approximated distribution of $W$ gives better test results. Thus, we use this bootstrap technique to compute the critical values for arbitrary significance levels.
\end{itemize}

\begin{table}[ht!]
	\centering
	\caption{Constructed covariance matrix $\hat{\mathbf{C}}^{H_0}(\bh,u)$ for different types of property in $H_0$.}
	\begin{tabular}{c|c}
		\hline
		\textbf{Type} & $\hat{\mathbf{C}}^{H_0}(\bh,u)$
		\\\hline
		
		$\text{V}^\text{sym}$
		& 
		$  \hat C_{ij}^{H_0}(\bh,u)  \defeq  \{\hat C_{ij}(\bh,u) + \hat C_{ji}(\bh,u)\}/2$
		\\\hline
		
		$\text{S}^\text{sym}$
		& 
		$  \hat C_{ij}^{H_0}(\bh,u)  \defeq  \{\hat C_{ij}(\bh,u) + \hat C_{ij}(-\bh,u)\}/2$
		\\\hline
		
		$\text{T}^\text{sym}$
		&
		$  \hat C_{ij}^{H_0}(\bh,u)  \defeq  \{\hat C_{ij}(\bh,u) + \hat C_{ij}(\bh,-u)\}/2$
		\\\hline
		
		V$\mid$ST& 
		$  \hat C_{ij}^{H_0}(\bh,u)  \defeq  \hat \rho_1(\bh,u)\hat C_{ij}(\bzero,0)$
		\\\hline
		
		S$\mid$VT& 
		$  \hat C_{ij}^{H_0}(\bh,u)  \defeq  \hat \rho_2(\bh)\hat C_{ij}(\bzero,u)$
		\\\hline
		
		T$\mid$VS& 
		$  \hat C_{ij}^{H_0}(\bh,u)  \defeq  \hat \rho_3(u)\hat C_{ij}(\bh,0)$
		\\\hline
		
		V$\mid$S& 
		$  \hat C_{ij}^{H_0}(\bh,u)  \defeq  \hat \rho_4(\bh,u)\hat C_{ij}(\bzero,u)$
		\\\hline
		
		V$\mid$T& 
		$  \hat C_{ij}^{H_0}(\bh,u)  \defeq  \hat \rho_5(\bh,u)\hat C_{ij}(\bh,0)$
		\\\hline
		
		S$\mid$T& 
		$  \hat C_{ij}^{H_0}(\bh,u)  \defeq  \hat \rho_{6,ij}(\bh)\hat C_{ij}(\bzero,u)$
		\\\hline
	\end{tabular}
	\label{tab:constructed-cov}
\end{table}

\begin{figure}[ht!]
	\centering
	\includegraphics[width=\textwidth]{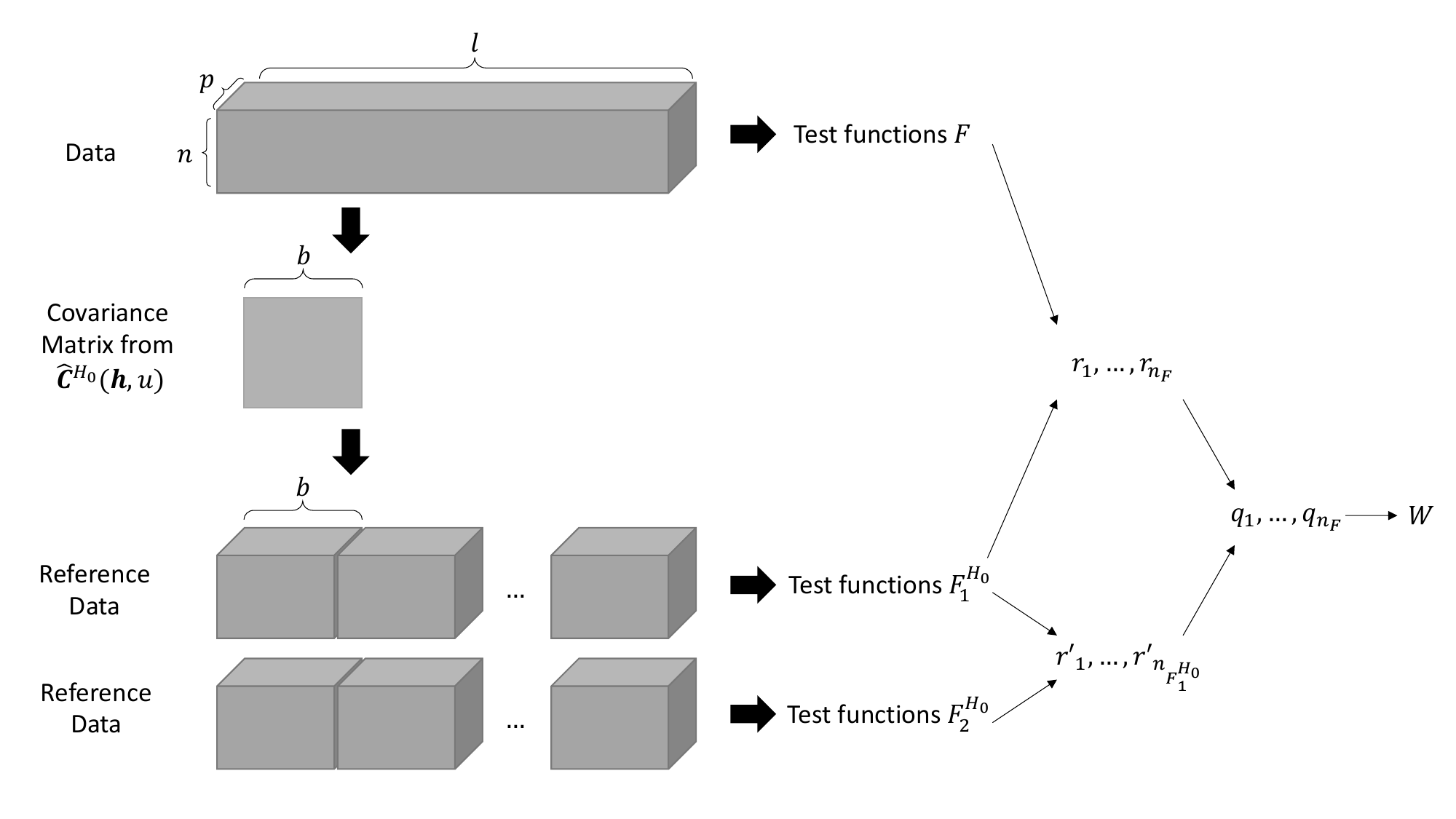}
	\caption{Flow chart of the proposed hypothesis testing procedure.}
	\label{fig:flowchart}
\end{figure}

The most challenging part of performing the test is to generate the reference data set in Step 3.
The dimension of the entire covariance matrix can be very large in the multivariate spatio-temporal case.
Memory and computational issues occur if we generate the reference data set as a whole. 
A more feasible approach is to generate the reference data block by block with a block size $b$ (temporal length) and assume that each block is only dependent on the previous block when $b$ is big. 
However, we find sensitivity issues for this approach in our study: a small error in the conditional distribution from the covariance sample estimators could make sequential conditional generation diverge.
To overcome this difficulty, we opt for a simpler approximation approach where all the blocks are generated independently.
This may lead to inconsistency for the generated reference data set to some extent. 
However, what we need from the reference data are the values of the test functions in temporal lag $u$, thus the covariance estimates of the simulated data for different $u$, rather than the generated data values themselves. 
In practice, the temporal lag $u$ considered for the test functions is kept small.
In fact, the errors in covariance estimates due to such an approximation only occur for time points around the boundary of the independent blocks, the number of which is negligible compared to the total sample pairs used to calculate the covariance estimate. 
Therefore, the obtained $F_1^{H_0}$ and $F_2^{H_0}$ can still reflect the correct variability as long as $b$ is not too small.
In general, when $b$ is larger, this artifact is more alleviated.
However, a larger $b$ causes more computation burden and memory consumption. 
We use a parameter $M$ to determine the size of the intermediate covariance matrix needed in the generation of reference data. 
There is a relationship between $M$ and the allowed maximum $b$:
$M\geq p\times n\times\min(b,l)$ (recall that $p$ is the number of variables, $n$ is the number of locations, and $l$ is the number of time points) for $\text{V}^\text{sym}$, $\text{S}^\text{sym}$, $\text{T}^\text{sym}$, V$\mid$S, V$\mid$T, and S$\mid$T; 
$M\geq\max(p,n\times\min(b,l))$ for V$\mid$ST because the needed covariance matrix can be written as a Kronecker product of two parts; 
$M\geq\max(n,p\times\min(b,l))$ for S$\mid$VT; 
and $M\geq\max(p\times n,\min(b,l))$ for T$\mid$VS.
One can set $M$ to the maximum value that is feasible for the computer in use. In our simulation study in Section~\ref{sec:sim}, we use $M=3000$, which is realistic for most laptops and desktops. We obtain results with good performance, and observe that in our simulation cases, larger values of $M$ do not lead to much different results but take much more computational time.
We illustrate the whole testing procedure in Figure~\ref{fig:flowchart}.
For non-Gaussian datasets, it is possible to extend the test procedure and apply the test to them if we know the distribution family so that the reference datasets can be simulated.

\section{Simulation Study}
\label{sec:sim}
\subsection{Visualization and assessment for symmetry property}\label{subsec:sym_sim}
To analyze the symmetry properties of multivariate spatio-temporal stationary covariances, we consider a bivariate spatio-temporal Gaussian random field $\bZ(\bs,t) = \{Z_1(\bs,t),Z_2(\bs,t)\}\trans$ for $l$ time points and $n=m^2$ locations in the unit square, i.e., $t \in \{1,2, \ldots, l\}$ and $\bs \in \{0,1/(m-1),\ldots,(m-2)/(m-1),1\}\times\{0,1/(m-1),\ldots,(m-2)/(m-1),1\}$. Model~\ref{model:sym} is used to generate data with different types of asymmetric covariance.

\begin{model}
\label{model:sym}
The second variable $Z_2(\bs,t)$ is a univariate first-order autoregressive spatio-temporal random field with a stationary isotropic Gaussian spatial noise. More specifically,
\[
\{Z_2(\bs_1,t),\ldots,Z_2(\bs_{n},t)\}\trans
=
\left\{
\begin{array}{ll}
0.5\{Z_2(\bs_1,t-1),\ldots,Z_2(\bs_{n},t-1)\}\trans + \boldsymbol{\varepsilon}_t&,t>1,\\
\boldsymbol{\varepsilon}_t&,t=1,\\
\end{array}
\right.
\]
where
$
\boldsymbol{\varepsilon}_1\sim N_n(\bzero,\bSigma)
$
and
$
\boldsymbol{\varepsilon}_t\sim N_n(\bzero,\frac{3}{4}\bSigma)
$
for $t>1$.
Here, $\bSigma$ is a matrix of dimension $n\times n$ with $(i,j)^\text{th}$ value $\Sigma_{ij} = \exp(-2\|\bs_i-\bs_j\|)$ for $i,j\in \{1,\ldots,n\}$.
The first variable $Z_1(\bs,t)$ is defined as
$Z_1(\bs,t)\defeq  \frac{\sqrt2}{2}Z_2(\bs+\Delta_\bs(\frac{1}{m-1},\frac{1}{m-1}),t+\Delta_t)+ \frac{\sqrt2}{2}\epsilon(\bs,t)$, where
$\epsilon(\bs,t)\sim N(0,1)$, $\Delta_t\geq0$ is the time lag, and $\Delta_\bs$ controls the distance of the spatial lag along the $45^\circ$ direction. To make $Z_1(\bs,t)$ well defined, $Z_2(\bs,t)$ is generated in a larger spatial grid and a longer time window as $\bs \in \{0,1/(m-1),\ldots,(m-1+\Delta_s)/(m-1)\}\times\{0,1/(m-1),\ldots,(m-1+\Delta_s)/(m-1)\}$ and $t\in\{1,2,\ldots,l+\Delta_t\}$. However, $Z_2(\bs,t)$ is eliminated when $\bs\notin [0,1]\times[0,1]$ or $t>l$ after obtaining all the needed $Z_1(\bs,t)$.

\end{model}
One can clearly see that in Model~\ref{model:sym}, $\Delta_s = \Delta_t = 0$ leads to a fully symmetric random field, $\Delta_s \neq 0$ leads to a random field that is not $\text{V}^\text{sym}$ or $\text{S}^\text{sym}$, and $\Delta_t \neq 0$ leads to a random field that is not $\text{V}^\text{sym}$ or $\text{T}^\text{sym}$.

\begin{figure}[ht]
	\centering
	\includegraphics[width=\textwidth]{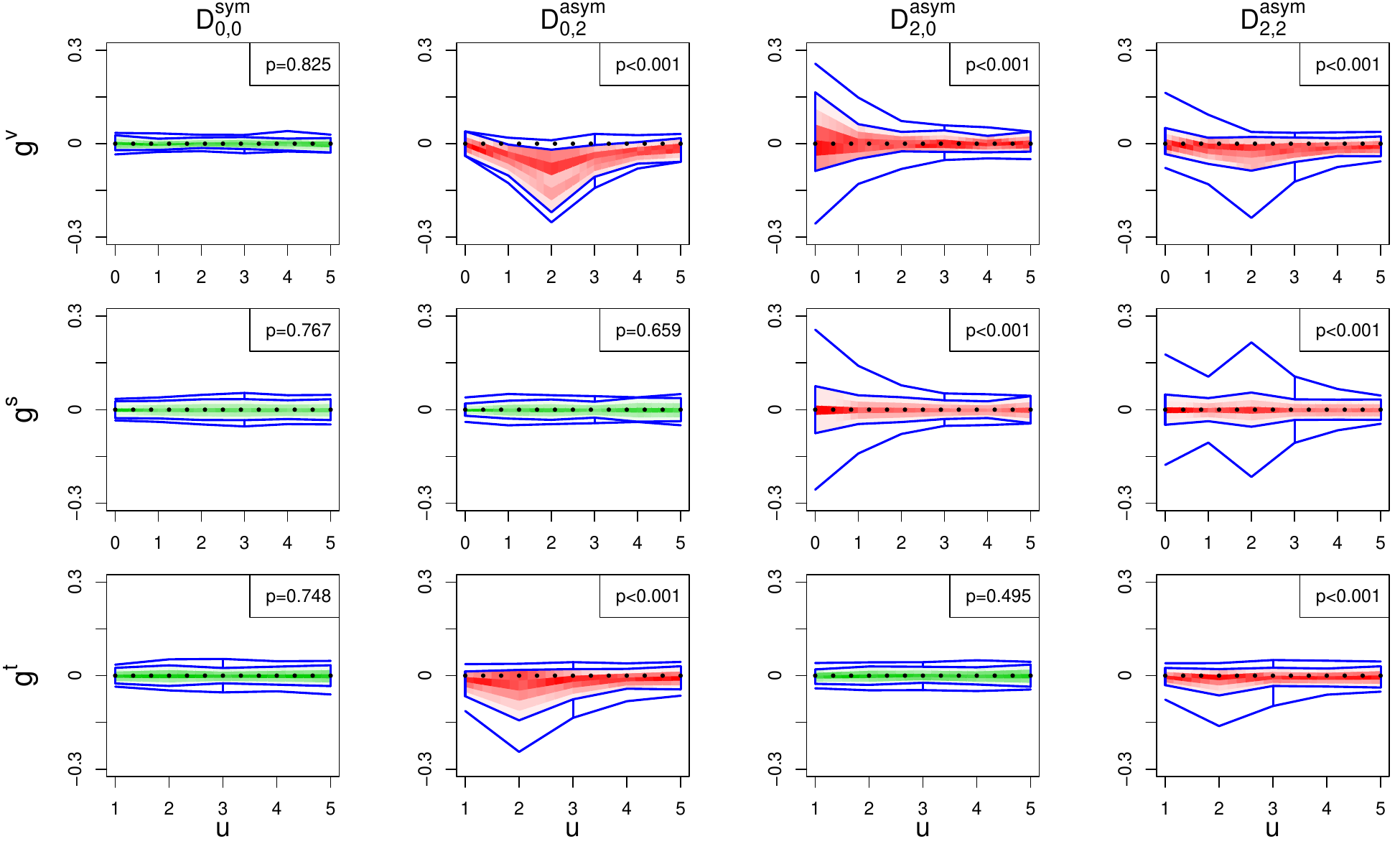}
	\caption{Visualization of symmetry test functions obtained from the simulated data $\text{D}_{0,0}^\text{sym}$, $\text{D}_{0,2}^\text{asym}$, $\text{D}_{2,0}^\text{asym}$, and $\text{D}_{2,2}^\text{asym}$.}
	\label{fig:sym}
\end{figure}

Figure~\ref{fig:sym} exhibits all the test functions $g^v$, $g^s$, and $g^t$ of one random data set generated from Model~\ref{model:sym} with $m=4$ and $l=10,000$. Four examples are shown with different combinations of chosen values of $\Delta_\bs$ and $\Delta_t$ (denoted by $\text{D}_{0,0}^\text{sym}$ when $\Delta_s = \Delta_t = 0$ and $\text{D}_{\Delta_\bs,\Delta_t}^\text{asym}$ when $\Delta_\bs\neq0$ or $\Delta_t\neq0$). From the visualization and the obtained p-values in the hypothesis testing, we observe the conclusions coinciding with the truth that when $\Delta_\bs\neq0$ or $\Delta_t\neq0$  the covariance is always not $\text{V}^\text{sym}$, and not $\text{S}^\text{sym}$ or $\text{T}^\text{sym}$ according to the non-zero $\Delta_s$ or $\Delta_t$. When both $\Delta_s$ and $\Delta_t$ are non-zero, the covariance does not satisfy any property of  $\text{V}^\text{sym}$, $\text{S}^\text{sym}$ and $\text{T}^\text{sym}$.

To show how the hypothesis testing performs in assessing the multivariate spatio-temporal symmetry property, we generate the four types of data $\text{D}_{0,0}^\text{sym}$, $\text{D}_{0,2}^\text{asym}$, $\text{D}_{2,0}^\text{asym}$, and $\text{D}_{2,2}^\text{asym}$ with 1000 replicates. We use a significance level of $5\%$ in the hypothesis testing, where 
$M$ is set as 3000 and $1000$ bootstraps are used. The results for the percentage of rejection replicates for each data type are given in Table~\ref{tab:sym-test}.
\begin{table}[ht!]
	\centering
	\caption{Percentage of rejections in 1000 replicates of data generated by Model \ref{model:sym} for each data type. All three types of symmetry properties are tested. Bold values are size, and others are power. Values in parentheses are estimated standard errors.  The significance level is $5\%$, $M=3000$, and $1000$ bootstrap samples are used in the hypothesis test.}
	\begin{tabular}{c||c|c||c|c}
		\hline
		Type & $\text{D}_{0,0}^\text{sym}$ & $\text{D}_{0,2}^\text{asym}$ &  $\text{D}_{2,0}^\text{asym}$ & $\text{D}_{2,2}^\text{asym}$ \\
		\hline
		\hline
		$\text{V}^\text{sym}$ &\textbf{5.1(0.7)} & 100.0(0.0) & 100.0(0.0) & 100.0(0.0) \\ \hline
		$\text{S}^\text{sym}$ &\textbf{6.0(0.8)} & \textbf{6.5(0.8)} & 100.0(0.0) & 100.0(0.0) \\ \hline
		$\text{T}^\text{sym}$ &\textbf{5.6(0.7)} & 100.0(0.0) & \textbf{5.3(0.7)} & 100.0(0.0) \\ \hline
	\end{tabular}
	\label{tab:sym-test}
\end{table}
All the bold values indicate the associated properties hold for the particular data set, meaning the size of the test. Since we use the significance level of $5\%$, the percentage of rejected cases should be ideally close to $5\%$. We see the size tends to be slightly higher than the significance level, but still within two standard errors. All the other cases are reflecting the power, where the associated properties do not hold. We see the proposed hypothesis test has a very high power, detecting asymmetric covariances in all the replicates.

\subsection{Visualization and assessment for separability property\label{subsec:test_sep}}
In this separability study, we consider a zero-mean trivariate spatio-temporal Gaussian random field $\bZ(\bs,t) = \{Z_1(\bs,t),Z_2(\bs,t),Z_3(\bs,t)\}\trans$ for $t \in \{1,2, \ldots, l\}$ and $\bs \in \{0,1/(m-1),\ldots,(m-2)/(m-1),1\}\times\{0,1/(m-1),\ldots,(m-2)/(m-1),1\}$. 
Following the way of building covariance models through latent dimensions by \citet{AG2010} or using products of nonseparable functions by \citet{G2002}, we use a valid covariance function as used in Model~\ref{model:sep}.
\begin{model}
	\label{model:sep}
	The trivariate Gaussian process $\bZ(\bs,t) = \{Z_1(\bs,t),Z_2(\bs,t),Z_3(\bs,t)\}\trans$ has mean zero and the following stationary covariance function:	
	$$
	C_{ij}(\bh,u) = 
	\dfrac{1}{(\abs{0.2u}+1)(\abs{i-j}+1)}
	\exp\left(-\dfrac{\abs{0.2u} ^2}{{(\abs{i-j}+1)}^{\beta_1}}
	-\dfrac{\|\bh\|^2}{{(\abs{0.2u}+1)}^{\beta_2}}\right), i,j=1,2,3.
	$$
\end{model}

\begin{figure}[ht!]
	\centering
	\includegraphics[width=\textwidth]{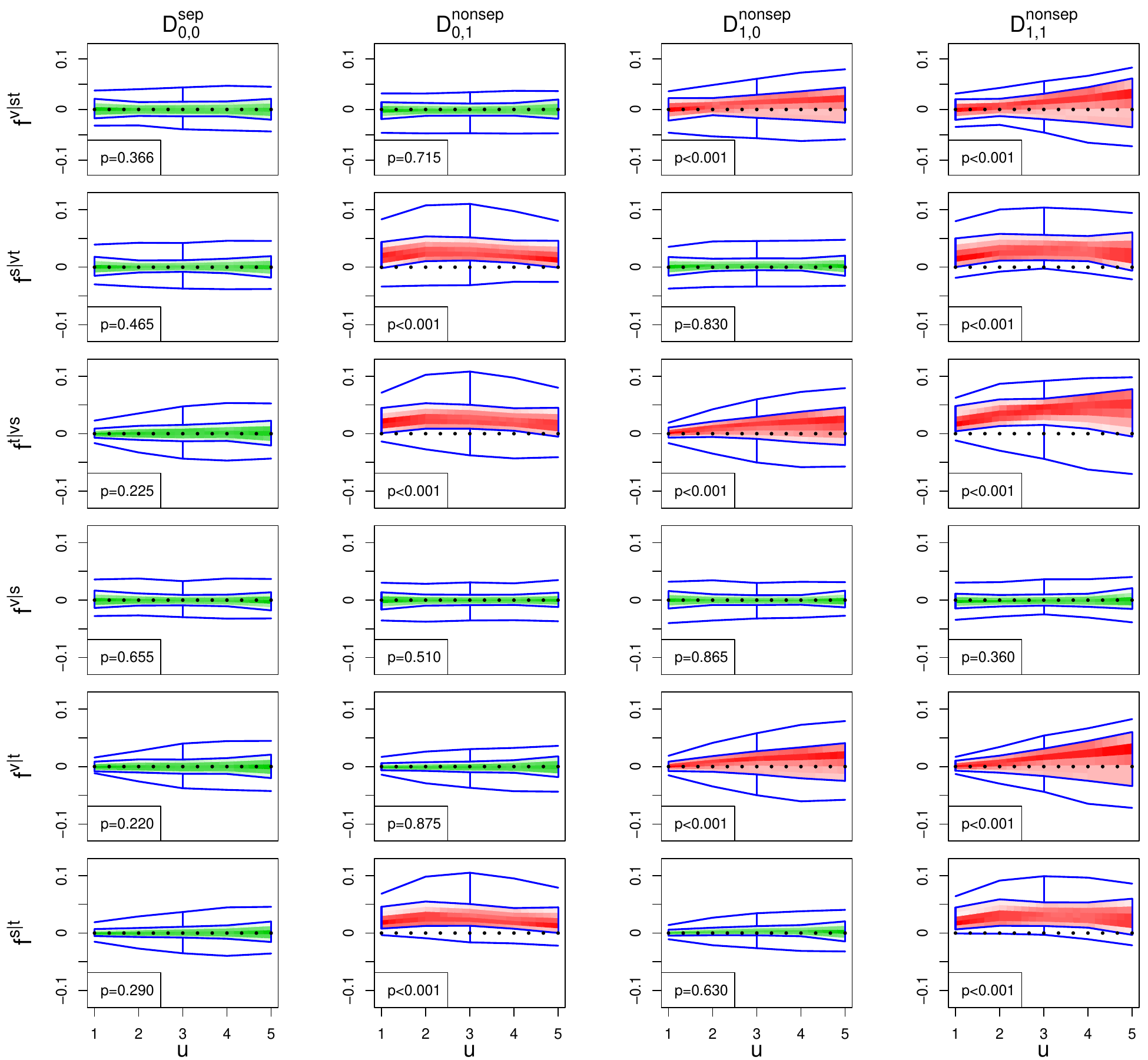}
	\caption{Visualization of separability test functions obtained from the simulated data $\text{D}_{0,0}^\text{sep}$, $\text{D}_{0,1}^\text{nonsep}$, $\text{D}_{1,0}^\text{nonsep}$, and $\text{D}_{1,1}^\text{nonsep}$.}
	\label{fig:sep}
\end{figure}

We can easily observe that, when $\beta_1=\beta_2=0$, the covariance is $\text{F}^\textup{sep}$; when $\beta_1=0,\beta_2\neq0$, the covariance is V$\mid$ST, V$\mid$S and V$\mid$T; when $\beta_1\neq0,\beta_2=0$, the covariance is S$\mid$VT, V$\mid$S and S$\mid$T;
and when $\beta_1\neq0,\beta_2\neq0$, the covariance is V$\mid$S.
Figure~\ref{fig:sep} exhibits all the test functions obtained from simulated data (denoted by $\text{D}_{0,0}^\text{sep}$ when $\beta_1 = \beta_2 = 0$ and denoted by $\text{D}^\text{nonsep}_{\beta_1,\beta_2}$ when $\beta_1\neq0$ or $\beta_2\neq0$) with $m=4$, $l=10,000$ for different values of $\beta_1$ and $\beta_2$. The visualization and obtained p-values reflect the correct covariance structure in theory. We also observe that the covariance built by Model~\ref{model:sep} is always V$\mid$S because we do not add a variables-space interaction term in the covariance model.

The size and power study of the hypothesis testing for the multivariate spatio-temporal separability property also uses the synthetic data from Model~\ref{model:sep} with different values of $\beta_1$ and $\beta_2$. To have more insights into the trend of power on non-zero $\beta_1$ or $\beta_2$ values, we use three distinct values $0,0.5,1$ for both $\beta_1$ and $\beta_2$.
We still generate 1000 replicates for each data type and use a significance level of $5\%$ in the test, where 
$M$ is set as 3000 and $200$ bootstraps are used.
The results for the number of rejection replicates for each data type are given in Table~\ref{tab:sep-test}, where the bold values are the size and others are power. The obtained sizes are close to the true nominal level $5\%$, and the power generally increases as $\beta_1$ or $\beta_2$ increases when more interaction is introduced. For the most extreme case when $\beta_1=\beta_2=1$, all the powers are above $95\%$. 

\begin{table}[ht!]
	\scriptsize
	\centering
	\caption{Percentage of rejection in 1000 replicates of data generated by Model \ref{model:sep} for each data type. All six types of separability properties are tested. Bold values are size, while others are power. Values in parentheses are estimated standard errors.  The significance level is $5\%$, $M=3000$, and $200$ bootstrap samples are used in the hypothesis testing.}
	\begin{tabular}{c||c|c|c||c|c|c||c|c|c}
		\hline
		\multirow{2}{*}{Type}& \multicolumn{3}{c||}{$\beta_1=0$}	&
		\multicolumn{3}{c||}{$\beta_1=0.5$} &  \multicolumn{3}{c}{$\beta_1=1$}\\
		\cline{2-10}
		&	$\beta_2=0$ & 	$\beta_2=0.5$ & 	$\beta_2=1$ & 
		$\beta_2=0$ & 	$\beta_2=0.5$ & 	$\beta_2=1$ & 
		$\beta_2=0$ & 	$\beta_2=0.5$ & 	$\beta_2=1$ \\
		\hline
		\hline

V $\mid$ ST & \textbf{6.1(0.8)} & \textbf{5.7(0.7)} & \textbf{4.7(0.7)} & 76.3(1.3) & 81.3(1.2) & 87.1(1.1) & 100.0(0.0) & 100.0(0.0) & 100.0(0.0) \\ \hline
S $\mid$ VT & \textbf{6.1(0.8)} & 99.7(0.2) & 100.0(0.0) & \textbf{4.6(0.7)} & 100.0(0.0) & 100.0(0.0) & \textbf{5.0(0.7)} & 100.0(0.0) & 100.0(0.0) \\ \hline
T $\mid$ VS & \textbf{5.3(0.7)} & 100.0(0.0) & 100.0(0.0) & 80.0(1.3) & 100.0(0.0) & 100.0(0.0) & 100.0(0.0) & 100.0(0.0) & 100.0(0.0) \\ \hline
V $\mid$ S & \textbf{7.9(0.9)} & \textbf{6.4(0.8)} & \textbf{5.7(0.7)} & \textbf{6.9(0.8)} & \textbf{6.9(0.8)} & \textbf{6.3(0.8)} & \textbf{8.0(0.9)} & \textbf{5.8(0.7)} & \textbf{6.1(0.8)} \\ \hline
V $\mid$ T & \textbf{5.6(0.7)} & \textbf{3.9(0.6)} & \textbf{4.7(0.7)} & 81.2(1.2) & 89.5(1.0) & 93.6(0.8) & 100.0(0.0) & 100.0(0.0) & 100.0(0.0) \\ \hline
S $\mid$ T &\textbf{6.1(0.8)} & 100.0(0.0) & 100.0(0.0) & \textbf{4.3(0.6)} & 100.0(0.0) & 100.0(0.0) & \textbf{4.6(0.7)} & 100.0(0.0) & 100.0(0.0) \\ \hline

\hline                                                                                             
	\end{tabular}
	\label{tab:sep-test}
	
\end{table}

We also give the hypothesis testing results when the mean-ratio estimator of the $\rho$-functions are used (see Section~\ref{subsec:tf}) in Table~\ref{tab:sep-test-mean-ratio} in the Supplementary Material. Most of the results are similar, but our least-square estimators show Type I errors much closer to the nominal level than the mean-ratio estimates for V$\mid$S.

\section{Application to Bivariate Wind Data}
\label{sec:app}
\begin{figure}[t!]
	\centering
	\includegraphics[width=\textwidth]{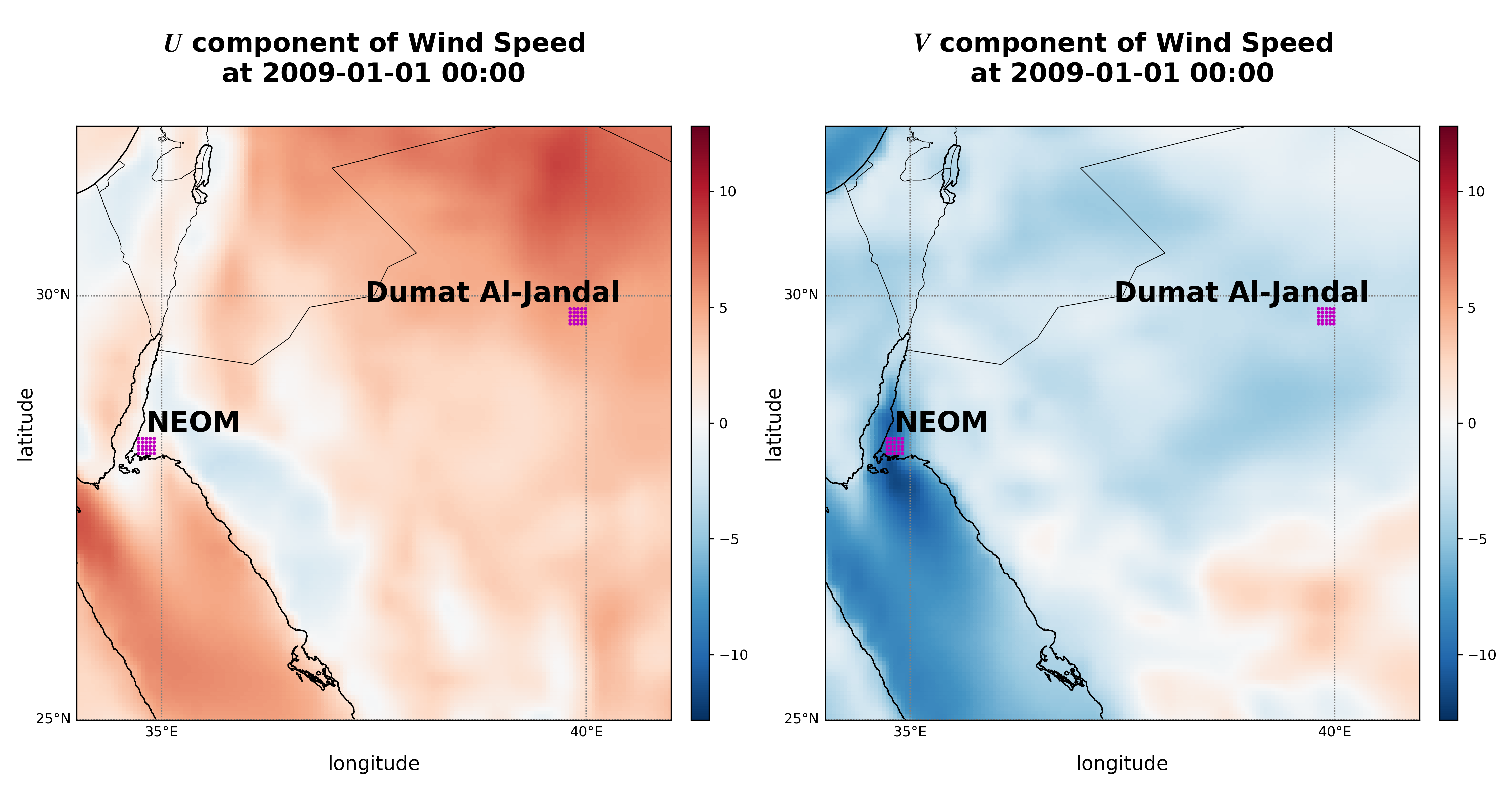}
	\caption{The bivariate wind speed at 00:00, January 1, 2009. A $5\times5$ grid in each of the two areas (NEOM city and Dumat Al-Jandal wind farm) are selected, shown as the magenta points.}
	\label{fig:wind_speed}
\end{figure}

In this section, we apply our visualization and assessment method to test the covariance structure in wind speed, which is a very important variable in many environmental studies. Wind farms and power grids are especially interested in obtaining sensible models of wind speed to better operate and manage the devices. For this purpose, we study the bivariate hourly wind speed in two areas in Saudi Arabia: one is an inland wind farm, Dumat Al-Jandal, currently being built; the other one is a new mega-city, NEOM, in the northwestern coast, which is still under construction and expected to consume a large amount of renewable energy (wind and solar). The two areas are shown in Figure~\ref{fig:wind_speed}, where we choose $5\times5=25$ locations in each area. We use high-resolution wind speed data in 2009 simulated by the Weather Forecasting and Research (WRF) model from \cite{Y2018}. The $U$ and $V$ components corresponding to two orthogonal directions of the wind vector are used as the bivariate variable. Figure~\ref{fig:wind_speed} depicts the $U$ and $V$ components of the wind speed at 00:00, January 1st, 2009.
After exploring the data set by Fourier transformation, we find strong periodic variability associated with 12-hour and 24-hour periods. Thus, we use a harmonic regression to remove the periodic mean and the intercept for each variable at each location as follows,
\[
\begin{array}{rcl}
X(\bs,t) &=&\beta_{Z,0}(\bs)  + \beta_{Z,1}(\bs)\cos(2\pi t/24)  + \beta_{Z,2}(\bs)\sin(2\pi t/24) \\
&&+\beta_{Z,3}(\bs)\cos(2\pi t/12)  + \beta_{Z,4}(\bs)\sin(2\pi t/12) + \tilde X(\bs,t), ~X\in\{U,V\}.\\
\end{array}
\]
After the regression, the remaining process $\bZ(\bs,t):=\{ \tilde U(\bs,t), \tilde V(\bs,t)\}\trans$ becomes zero-mean, and we assess its covariance structure.

Wind has a very complex dynamic, and the structure may change from time to time~\citep{VGPM2010}. 
We analyze the bivariate hourly wind speed $\bZ(\bs,t)$ for each month in 2009 and assume $\bZ(\bs,t)$ is stationary for each month as an example to show the intra-annual variability of the wind structure.
As revealed in previous studies, wind speed has often a strong interaction between space and time, and the prevailing wind direction generally makes the wind speed asymmetric. For this reason, we choose a less conservative significance level, $10\%$, to perform the hypothesis test.

\begin{figure}[ht!]
	\centering
	\includegraphics[width=\textwidth]{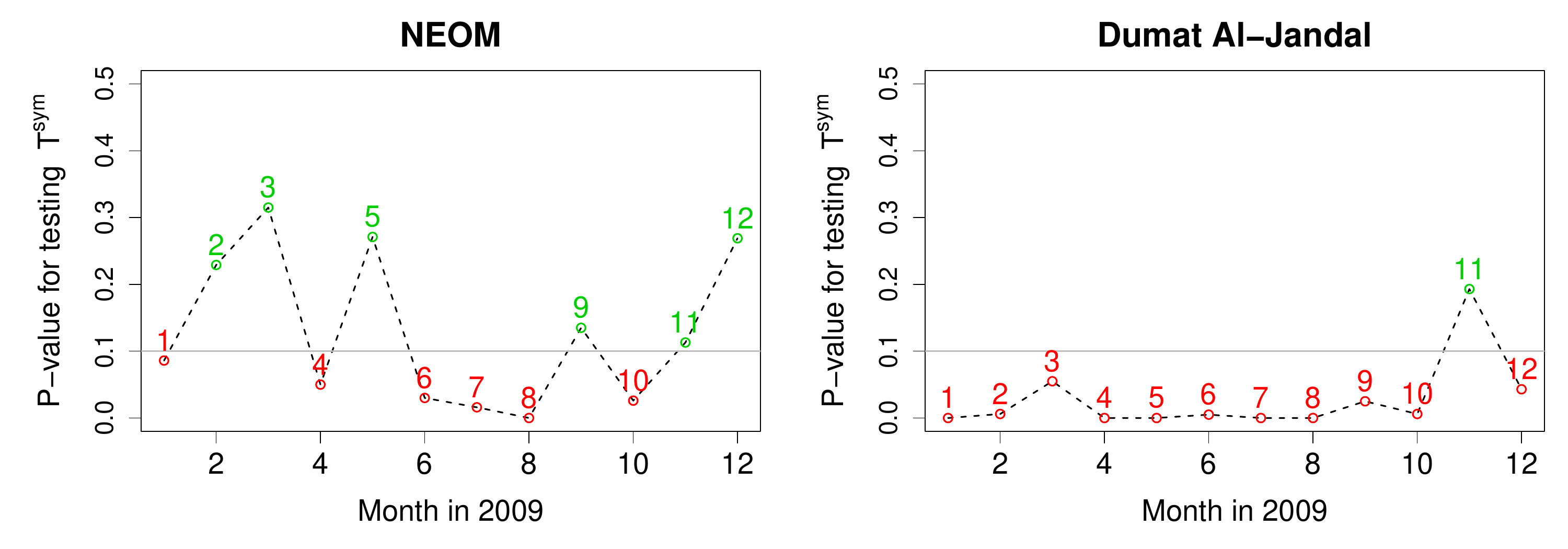}
	\caption{P-values of testing $\text{T}^\text{sym}$ for the covariance in each month in NEOM and Dumat Al-Jandal. 
	We show the month numbers in red if $\text{T}^\text{sym}$ is rejected and in green otherwise when using the significance level $10\%$.}
	\label{fig:app_sym_res}
\end{figure}

\begin{figure}[ht!]
	\centering
	\includegraphics[width=\textwidth]{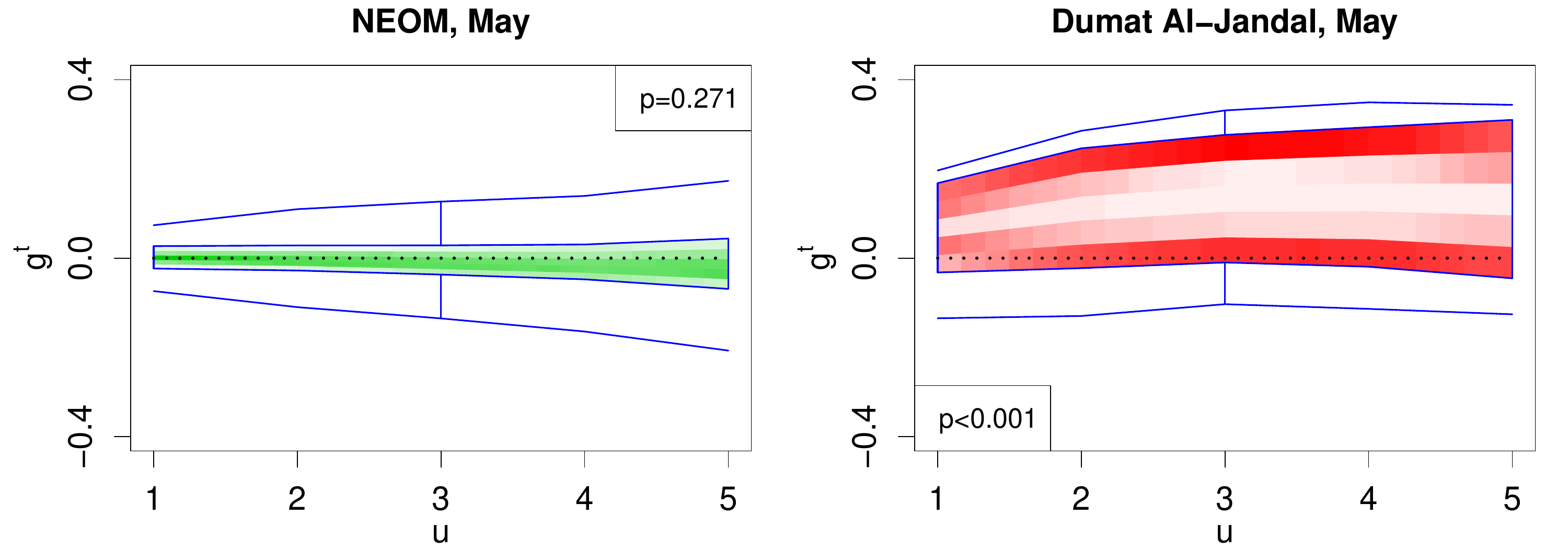}
	\caption{Visualization of test functions $g^t$ in May at NEOM and Dumat Al-Jandal. The green plot shows $g^t$ where $\text{T}^\text{sym}$ is not rejected, and the red plot shows $g^t$ where $\text{T}^\text{sym}$ is rejected when using the significance level $10\%$.
	}
	\label{fig:app_sym_res_example}
\end{figure}

For the symmetry test, we find that $\text{V}^\text{sym}$ and $\text{S}^\text{sym}$  are rejected in all cases. Figure~\ref{fig:app_sym_res} summarizes the p-values of testing $\text{T}^\text{sym}$ for the two areas in each month. 
$\text{T}^\text{sym}$ is not rejected for more months in NEOM, while only November in Dumat Al-Jandal shows $\text{T}^\text{sym}$. 
We also observe that the covariance for all summer months (June, July, August) are not $\text{T}^\text{sym}$, inferring a more complex structure of wind in summer, due to potentially prevailing wind direction. Figure~\ref{fig:app_sym_res_example} gives the visualized test functions $g^t$ for May in the two areas.

\begin{figure}[ht!]
	\centering
	\includegraphics[width=\textwidth]{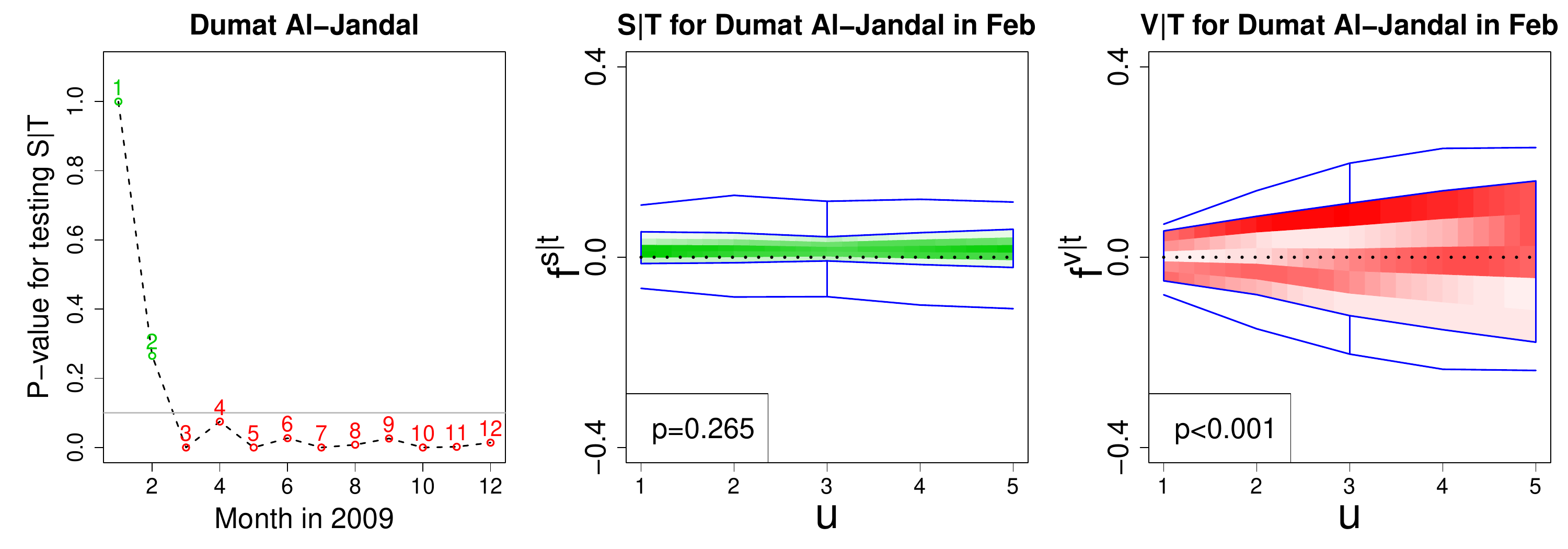}
	\caption{
		P-values of testing S$\mid$T for the covariance in each month in  Dumat Al-Jandal (left).
		We show the month numbers in red if the corresponding property S$\mid$T is rejected and in green otherwise when using the significance level $10\%$.
		Visualization of test functions $f^{s\mid t}$ (middle) and $f^{v\mid t}$ (right) for February in Dumat Al-Jandal, with red and green indicating whether the corresponding separability property is rejected using the significance level $10\%$.
	}
	\label{fig:app_sep_res}
\end{figure}

All the rejected symmetry assumptions lead to the rejection of the corresponding separability assumptions of V$\mid$ST, S$\mid$VT, and T$\mid$VS. 
We examined the rest and found that only S$\mid$T is not rejected in Dumat Al-Jandal in January and February, the p-values for which are given in Figure~\ref{fig:app_sep_res}. 
The fact that no separability properties in the first category (V$\mid$ST, S$\mid$VT, and T$\mid$VS) in any month and area are observed implies the general inappropriateness to use a Kronecker product covariance model in analyzing wind speed. 
Figure~\ref{fig:app_sep_res} also gives the visualized test functions $f^{s\mid t}$ and $f^{v\mid t}$ for February in Dumat Al-Jandal.
It is noteworthy that January and February in Dumat Al-Jandal do not show V$\mid$T; 
this is not surprising, because V$\mid$T and S$\mid$T lead to T$\mid$VS, and subsequently $\text{T}^\text{sym}$ by Proposition~\ref{prop:sep2}, but we know $\text{T}^\text{sym}$ is rejected for the two months.

\section{Discussion}
\label{sec:discussion}
In this work, we elaborated on different types of symmetry and separability properties of multivariate spatio-temporal stationary covariances.
We developed test functions associated with each property to visualize and assess them.
We used and modified the functional boxplot to visualize the developed test functions so that insights into the underlying covariance structures can be obtained.
We proposed a rank-based testing procedure to examine these properties in a more formal way, with demonstrated good size and high power in the simulation study.
We applied these tools to study the covariance of the bivariate wind speed in two areas of Saudi Arabia.

Obtaining and visualizing the test functions are always very fast.
When the deviation of test functions from zero is clearly observed, one may directly proceed with covariance models that do not assume the corresponding simplified property, such as the test functions for S$\mid$VT shown in Figure~\ref{fig:tfplot_example} (B), where we see the entire central region is above zero.
However, when a visual inspection cannot find obvious evidence to reject the null hypothesis, the testing procedure is then needed to provide a better indication by p-values, which is generally slower and needs more computational time due to boostrap and the generation of data under the null hypothesis.

Our proposed testing procedure also has limitations. As we showed in Section~\ref{sec:properties}, there are some constraints on the covariance properties.
What we proposed is an independent individual testing scheme. In practice, there may be contradictory testing results from different independent tests. Developing a multi-testing framework may potentially resolve this problem. However, rigorous design and careful power studies are needed. This would be a direction for future research.

It is also noteworthy that what we have studied is the overall property for the multivariate spatio-temporal covariance. The proposed symmetry or separability requires the corresponding covariance property to hold for every pair of variables. It is possible to define more subtypes from various emphasized perspectives where only a part of the variables is required to meet the requirement. 
However, these subtypes are rather trivial extensions, and the developed test can be easily adapted to them.

The way we study multivariate spatio-temporal covariance properties is by building univariate test functions.
Alternatives to our proposed approach are also possible, such as developing multivariate test functions and then visualizing and testing them based on multivariate functional data depths~\citep[e.g.,][]{LSLG2014}.
The performance of this work can be used as a benchmark for future techniques to be developed.

The \textsf{R} code~\citep{R2021} for our proposed visualization and test methods (code.zip) is provided in the Supplementary Material. In addition, an interactive \textsf{R} ShinyApp (shiny.zip) is also provided in the Supplementary Material and available at \href{https://hhuang.shinyapps.io/mstCovariance}{\url{https://hhuang.shinyapps.io/mstCovariance}}, where one can easily make different settings in the simulation examples described in Section~\ref{sec:sim} and see how the multivariate space-time covariance properties are changed.

\section*{\centering Acknowledgments}
This publication is based on research supported by the King Abdullah University of Science and Technology (KAUST) Office of Sponsored Research (OSR) under Award No: OSR-2018-CRG7-3742 and in part by the Center of Excellence for NEOM Research at KAUST. 
The authors report that there are no competing interests to declare.

\section*{Supplementary Material}

\begin{description}
	\item[supplementary-document.pdf:] Proofs and supplementary table
	\item[code.zip:] \textsf{R} code for our proposed visualization and test methods
	\item[shiny.zip:] Interactive \textsf{R} ShinyApp showing simulation examples
\end{description}

\bibliographystyle{apalike}
\bibliography{reference}

\clearpage

\setcounter{equation}{0}
\renewcommand{\theequation}{S\arabic{equation}}

\setcounter{figure}{0}
\setcounter{table}{0}
\setcounter{section}{0}
\setcounter{page}{1}
\renewcommand{\thesection}{S\arabic{section}}
\renewcommand{\thefigure}{S\arabic{figure}}
\renewcommand{\thetable}{S\arabic{table}}

\begin{center}
	\bf\large  Supplementary Document to
			``Test and Visualization of Covariance Properties for Multivariate Spatio-Temporal Random Fields''
			published in the Journal of Computational and Graphical Statistics

\end{center}

\begin{center}
	Huang Huang,
	Ying Sun, and
	Marc G. Genton\\
Statistics Program\\
King Abdullah University of Science and Technology\\

\today
\end{center}

\section{Proofs}

\begin{proof}[\textup{\textbf{Proof of Proposition~\ref{prop:sym}:}} ]
We assume the covariance is \textit{symmetric in space} and \textit{symmetric in time}. Then, for any space lag $\bh$ and time lag $u$, we have $C_{ij}(\bh,u) = C_{ij}(-\bh,u)$ by \textit{symmetry in space} and $C_{ij}(-\bh,u) = C_{ij}(-\bh,-u)$ by \textit{symmetry in time}, for any $i,j=1,\ldots,p$. Therefore, $C_{ij}(\bh,u) = C_{ij}(-\bh,-u), \forall i,j\in 1,\ldots,p$, which makes the covariance \textit{symmetric in variables}. The proof is similar for other situations.
\end{proof}
\begin{proof}[\textup{\textbf{Proof of Proposition~\ref{prop:sep1}:}} ]
We assume the covariance is V$\mid$ST. Then, we know $C_{ij}(\bh,u) =  \rho_1(\bh,u)C_{ij}(\bzero,0)$. Plugging in $\bh=\bzero$, we get $C_{ij}(\bzero,u) =  \rho_1(\bzero,u)C_{ij}(\bzero,0)$. Thus, we have $C_{ij}(\bh,u) =  \rho_1(\bh,u) C_{ij}(\bzero,u)/  \rho_1(\bzero,u)$. Let $\rho_4(\bh,u):=\rho_1(\bh,u)/\rho_1(\bzero,u)$, so we know $\rho_4(\bzero,u)=1,\forall u$ and $C_{ij}(\bh,u) =  \rho_4(\bh,u) C_{ij}(\bzero,u)$, which is V$\mid$S. Similarly, plug in $u=0$ for the equation $C_{ij}(\bh,u) =  \rho_1(\bh,u)C_{ij}(\bzero,0)$ and we get $C_{ij}(\bh,0) =  \rho_1(\bh,0)C_{ij}(\bzero,0)$. Thus, $C_{ij}(\bh,u) =  \rho_1(\bh,u) C_{ij}(\bh,0)/  \rho_1(\bh,0)$. Let $\rho_5(\bh,u):=\rho_1(\bh,u)/\rho_1(\bh,0)$. Then, we have $\rho_5(\bh,u)=1,\forall \bh$ and $C_{ij}(\bh,u) =  \rho_5(\bh,u) C_{ij}(\bzero,u)$, which is V$\mid$T. The proof is similar for other situations.
\end{proof}
\begin{proof}[\textup{\textbf{Proof of Proposition~\ref{prop:sep2}:}} ]
We assume the covariance is V$\mid$S and V$\mid$T. Then,
$C_{ij}(\bh,u) = \rho_4(\bh,u)C_{ij}(\bzero,u)$ and $C_{ij}(\bh,u) = \rho_5(\bh,u)C_{ij}(\bh,0)$, so we get the equation $\rho_4(\bh,u)C_{ij}(\bzero,u) = \rho_5(\bh,u)C_{ij}(\bh,0)$. Plugging in $\bh=\bzero$, we have $C_{ij}(\bzero,u) = \rho_5(\bzero,u)C_{ij}(\bh,0)/\rho_4(\bzero,u)$. As $\rho_4(\bzero,u)=1$, we know $C_{ij}(\bzero,u) = \rho_5(\bzero,u)C_{ij}(\bh,0)$. Therefore, we obtain $C_{ij}(\bh,u) = \rho_4(\bh,u)\rho_5(\bzero,u)C_{ij}(\bzero,0)$. Let $\rho_1(\bh,u)\defeq\rho_4(\bh,u)\rho_5(\bzero,u)$, and we have $\rho_1(\bzero,0)=\rho_4(\bzero,0)$ $\rho_5(\bzero,0)=1$. Eventually, we get $C_{ij}(\bh,u) = \rho_1(\bh,u)C_{ij}(\bzero,0)$, which is V$\mid$ST. The proof is similar for other situations.
\end{proof}
\begin{proof}[\textup{\textbf{Proof of Proposition~\ref{prop:sep3}:}} ]
We assume the covariance is V$\mid$ST and S$\mid$VT. By Proposition~\ref{prop:sep1}, we know the covariance is also V$\mid$T (as it is V$\mid$ST) and S$\mid$T (as it is S$\mid$VT). Then, by Proposition~\ref{prop:sep2}, the covariance is also T$\mid$VS. Moreover, V$\mid$S also holds from the fact that the covariance is V$\mid$ST. Therefore, all the separability properties hold, and the covariance is F$^\text{sep}$.
\end{proof}

\begin{proof}[\textup{\textbf{Proof of Theorem~\ref{th:sep}:}} ]
We only prove the case for test functions of separability between variables and space-time, and the proof is similar for other separability cases.
For any finite $u\in\mathbb{N}_+$ and $\bs_a, \bs_b$,
we define
$
\bc=
\big(
C_{11}(\bzero,0),C_{12}(\bzero,0)\cdots,C_{pp}(\bzero,0),
C_{11}(\bs_b-\bs_a,u),C_{12}(\bs_b-\bs_a,u)\cdots,C_{pp}(\bs_b-\bs_a,u)
\big)^\top
$
and denote by $\hat\bc$ the sample estimate of $\bc$ whose entries have the form
$  
\hat C^{a,b}_{ij}(\bs_b-\bs_a,u) \defeq \dfrac{1}{l-u}
\sum^{l-u}_{t=1}
\Bigg\{Z_j(\bs_b,t+u)-\frac{\sum\limits^{l-u}_{r=1}Z_j(\bs_b,r+u)}{l-u}\Bigg\}
\Bigg\{Z_i(\bs_a,t)-\frac{\sum\limits^{l-u}_{r=1}Z_i(\bs_a,r)}{l-u}\Bigg\}.
$
Proposition 1 in \citet{LGS2008} shows that under the condition
$\sum\limits_{t\in\mathbb{Z}}
\abs{\cov\{Z_i(\bs_a,0)Z_{j}(\bs_b,u_1), \allowbreak Z_{i'}(\bs_{a'},t)Z_{j'}(\bs_{b'},t+u_2)\}}<\infty$,
for any finite $u_1,u_2\in\mathbb{Z}$, $i, j, i', j' \in\{1,\ldots,p\}$, and $\bs_a,\bs_b,\bs_{a'},\bs_{b'}\in\mathcal{D}$,
we have
$
l^{1/2}(\bc-\hat\bc)\rightarrow N_{2p^2}(\boldsymbol{0},\bSigma)
$
in distribution as $l\rightarrow\infty$, 
where $\bSigma = l\lim\limits_{l\rightarrow\infty}\cov(\hat\bc,\hat\bc)$.
Applying the multivariate delta theorem~\citep{BMK1979},
we have
$
l^{1/2}\big(J(\bc)-J(\hat\bc)\big)\rightarrow N\big(\boldsymbol{0},\nabla_\bc J(\bc)^\top\bSigma\nabla_\bc J(\bc)\big)
$
in distribution as $l\rightarrow\infty$ for any function $J(\cdot)$ differentiable at $\bc$.

For any $i,j\in\{1,\ldots,p\}$, we let
\[
J(\bc;i,j) = C_{ij}(\bzero,0)\dfrac{\sum\limits_{i',j'=1}^pC_{i'j'}(\bs_b-\bs_a,u)C_{i'j'}(\bzero,0)}{\sum\limits_{i',j'=1}^pC^2_{i'j'}(\bzero,0)},
\]
and we know $J(\bc;i,j)$ is differentiable at $\bc$ when $\sum\limits^{p}_{i',j'=1}C^2_{i'j'}(\bzero,0)\neq 0$.
Under the null hypothesis (separability between variables and space-time), 
we have $C_{i'j'}(\bh,u) = \rho_1(\bs_b-\bs_a,u)C_{i'j'}(\bzero,0), \forall i',j'\in\{1,\ldots,p\}$. Then, 
\[
J(\bc;i,j) = C_{ij}(\bzero,0)
\dfrac{\sum\limits_{i',j'=1}^p\rho_1(\bs_b-\bs_a,u)C^2_{i'j'}(\bzero,0)}
{\sum\limits_{i',j'=1}^pC^2_{i'j'}(\bzero,0)} 
=
C_{ij}(\bzero,0)\rho_1(\bs_b-\bs_a,u)
=
C_{ij}(\bs_b-\bs_a,u).
\]
Therefore, we have 
\begin{align*}
\mathbb{E}[f^{v\mid st}_{i,j,a,b}(u)]  &~ =
\mathbb{E}[\hat C^{a,b}_{ij}(\bs_b-\bs_a,u) - \hat \rho^{a,b}_1(\bs_b-\bs_a,u)\{\hat C^{a,a}_{ij}(\bzero,0)+\hat C^{b,b}_{ij}(\bzero,0)\}/2] \\
&~= C_{ij}(\bs_b-\bs_a,u)\\
&~~~~~~
- \mathbb{E}\left[\{\hat C^{a,a}_{ij}(\bzero,0)+\hat C^{b,b}_{ij}(\bzero,0)\}
 \dfrac{\sum\limits^{p}_{i',j'=1}\hat C^{a,b}_{i'j'}(\bs_b-\bs_a,u)\{\hat C^{a,a}_{i'j'}(\bzero,0)+\hat C^{b,b}_{i'j'}(\bzero,0)\}}
{\sum\limits^{p}_{i',j'=1}\{\hat C^{a,a}_{i'j'}(\bzero,0)+\hat C^{b,b}_{i'j'}(\bzero,0)\}^2}\right] \\
& ~=  J(\bc;i,j)- \mathbb{E}[J(\hat\bc;i,j)] 
=  \mathbb{E}[J(\bc;i,j)- J(\hat\bc;i,j)]
\rightarrow 0,
\end{align*}
as $l\rightarrow 0$. 
The derivation holds for any $i,j\in\{1,\ldots,p\}$, $u\in\mathbb{N}_+$, and $\bs_a,\bs_b\in\mathcal{D}$.
Thus, we obtain the conclusion that the expectation of all the test functions $f^{v\mid st}_{i,j,a,b}(u)$ converges to zero as the number of time points goes to infinity when the covariance function is separable with the corresponding type.
\end{proof}

\clearpage
\section{Supplementary Table}
\begin{table}[ht!]
	\scriptsize
	\centering
	\caption{Percentage of rejection in 1000 replicates of data generated by Model \ref{model:sep} for each data type. The $\rho$-functions in building the test functions are estimated by the mean-ratio estimator. All six types of separability properties are tested. Bold values are size, while others are power. Values in parentheses are estimated standard errors.  The significance level is $5\%$, $M=3000$, and $200$ bootstrap samples are used in the hypothesis testing.}
	\begin{tabular}{c||c|c|c||c|c|c||c|c|c}
		\hline
		\multirow{2}{*}{Type}& \multicolumn{3}{c||}{$\beta_1=0$}	&
		\multicolumn{3}{c||}{$\beta_1=0.5$} &  \multicolumn{3}{c}{$\beta_1=1$}\\
		\cline{2-10}
		&	$\beta_2=0$ & 	$\beta_2=0.5$ & 	$\beta_2=1$ & 
		$\beta_2=0$ & 	$\beta_2=0.5$ & 	$\beta_2=1$ & 
		$\beta_2=0$ & 	$\beta_2=0.5$ & 	$\beta_2=1$ \\
		\hline
		\hline
		
V $\mid$ ST & \textbf{5.8(0.7)} & \textbf{5.2(0.7)} & \textbf{5.2(0.7)} & 72.6(1.4) & 79.9(1.3) & 85.2(1.1) & 100.0(0.0) & 100.0(0.0) & 100.0(0.0) \\ \hline
S $\mid$ VT & \textbf{6.7(0.8)} & 99.4(0.2) & 100.0(0.0) & \textbf{3.6(0.6)} & 99.9(0.1) & 100.0(0.0) & \textbf{4.6(0.7)} & 100.0(0.0) & 100.0(0.0) \\ \hline
T $\mid$ VS & \textbf{5.2(0.7)} & 100.0(0.0) & 100.0(0.0) & 79.3(1.3) & 100.0(0.0) & 100.0(0.0) & 100.0(0.0) & 100.0(0.0) & 100.0.0(0.0) \\ \hline
V $\mid$ S & \textbf{3.0(0.5)} & \textbf{0.3(0.2)} & \textbf{0.1(0.1)} & \textbf{0.8(0.3)} & \textbf{0.5(0.2)} & \textbf{0.0(0.0)} & \textbf{1.2(0.3)} & \textbf{0.3(0.2)} & \textbf{0.0(0.0)} \\ \hline
V $\mid$ T & \textbf{5.9(0.7)} & \textbf{4.1(0.6)} & \textbf{4.7(0.7)} & 83.0(1.2) & 89.7(1.0) & 94.2(0.7) & 100.0(0.0) & 100.0(0.0) & 100.0(0.0) \\ \hline
S $\mid$ T &\textbf{5.6(0.7)} & 100.0(0.0) & 100.0(0.0) & \textbf{4.4(0.6)} & 100.0(0.0) & 100.0(0.0) & \textbf{4.6(0.7)} & 100.0(0.0) & 100.0(0.0) \\ \hline
		
		\hline                                                                                             
	\end{tabular}
	\label{tab:sep-test-mean-ratio}
	
\end{table}

\end{document}